\documentstyle[twocolumn,aps,psfig]{revtex}
\begin{document}
 
\twocolumn[\hsize\textwidth\columnwidth\hsize\csname@twocolumnfalse\endcsname
\draft
\tighten
 
\title{Reply on ``Aging Effects in a Lennard-Jones Glass''}
 
\author{Walter Kob }
\address{ Institut f\"ur Physik, Johannes Gutenberg-Universit\"at,
Staudinger Weg 7, D-55099 Mainz, Germany}
 
\author{Jean-Louis Barrat }
\address{ D\'epartement de Physique des Mat\'eriaux \\
Universit\'e Claude Bernard and CNRS, 69622 Villeurbanne Cedex, France}

\date{April 6, 1998}
 
\maketitle

 
\vskip2pc]
In their comment to our Letter on aging in a Lennard-Jones
glass~\cite{letter}, M\"ussel and Rieger~\cite{comment} give evidence
that a different type of scaling function, specified below, might give
rise to a better scaling of the two-time correlation functions
$C_q(t_w,t_w+t)$ than the scaling function propsed by us. By using data
stemming from their own simulation, they show that a scaling of the
form
\begin{equation}
C_q(t_w,t_w+t)\sim \tilde{C}\{\ln((t+t_w)/\tau)/\ln(t_w/\tau)\},
\end{equation}
where $\tau$ is a fit parameter, leads to a better collapse of the
curves for the different waiting times $t_w$, than the scaling used by
us, which is of the form $C_q(t_w,t_w+t)\sim \tilde{C} \{t/t_r\}$,
with $t_r \propto t_w^{\alpha}$, $\alpha\approx 0.9$, and which thus
almost corresponds to ``simple aging'' ($t_r\propto t_w$).
We therefore used the scaling function proposed by M\"ussel and Rieger
in order to see whether this function can be used to scale also the
data from {\it our} simulation onto a master function and show the result in
Fig~1. The value of $\tau$ is 0.005, which is comparable to the one
found by M\"ussel and Rieger for their data. As is evident from the
figure, this sort of scaling (main figure) does not work well for our
data and is clearly inferior than the scaling proposed by us (inset).
Thus we conclude that the scaling function proposed by M\"ussel and
Rieger is not always appropriate to scale the two-time correlation
functions for this sort of system.

What remains unclear for the moment is the reason why the scaling
function given by Eq.~(1) works for the data of M\"ussel and Rieger,
whereas it fails to do so for our data. A careful comparison of their
and our data shows that the two sets of curves show some systematic
differences, in that, e.g, the height of the plateau at intermediate
times is slightly larger in their data than in our data. Also, as stated
by M\"ussel and Rieger, the $t_w$ dependence of the relaxation time
$t_r$ differs from ours, since they find that the exponent
$\alpha$ is 1.1 (as oposed to 0.88). The reason for the difference of 
the relaxation data
might be that the two simulations were not carried out in exactly the
same way.  For example the size of the system is different (32768 vs.
1000), the time step is different (0.01 vs. 0.02) and the value of the
wave-vector is different (7.5 vs. 7.25). It is not clear which ones of
these differences, if any, gives rise to the slightly different aging
behavior. It has been found before, however, that the details of the
nonequilibrium dyamics can depend quite sensitively on the details of
the simulations~\cite{letter,parisi}.  Finally we mention that if
$T_f$, the final temperature of the quench, is decreased significantly,
the scaling behavior found in Ref.~\cite{letter} does not seem to hold
anymore and that also the Ansatz given by Eq.~(1) does not seem to
work~\cite{barrat}

%\section*{Figures}
\begin{figure}[f]
\psfig{file=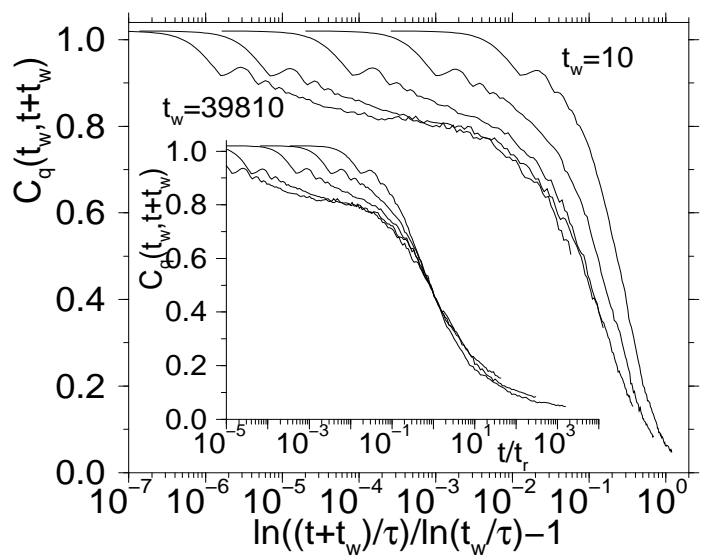,width=10cm,height=8cm}
\caption{Main figure: The data of Ref.~[1] scaled in the way proposed by M\"ussel and Rieger with
$\tau=0.005$. Note that the data for $t_w=0$ is not shown. 
Inset: The same data but scaled by $t_r\propto t_w^{0.88}$.}
\end{figure}
\end{document}